\documentclass[12pt]{article}
\usepackage{amsmath,amsthm,amssymb,amscd,fancybox,epsfig,float,subfigure}
\usepackage[margin=1in]{geometry}
\usepackage{authblk}
\usepackage{graphics}
\usepackage{epstopdf}
\usepackage{multirow}
\usepackage{hhline}
\usepackage[capitalise]{cleveref}
\usepackage{xcolor}
\usepackage{tikz}
\usepackage{mathtools}
\usetikzlibrary{arrows,automata}
\usepackage[nocompress,sort]{cite}

\newtheorem{remark}{Remark}[section]

\newtheorem{corollary}{Corollary}[section]

\newcommand{\beqn}{\begin{equation}}
\newcommand{\eeqn}{\end{equation}}

\newcommand{\cA}{\mathcal{A}}

\newcommand{\bR}{\mathbb{R}}

\newcommand{\bk}{{\boldsymbol k}}

\newcommand{\bx}{\boldsymbol{x}}

\newcommand{\by}{\boldsymbol{y}}
\newcommand{\cL}{\mathcal{L}}

\newcommand{\bbmat}{\begin{bmatrix}}
\newcommand{\ebmat}{\end{bmatrix}}
\newcommand{\rmax}{r_{\textrm{max}}}

\newtheorem{theorem}{Theorem}

\newtheorem{lemma}{Lemma}

\newtheorem{rem}{Remark}

\title{Analysis of single-excitation states in quantum optics}

\author[1]{Jeremy Hoskins}

\author[2,3]{Jason Kaye}

\author[2]{Manas Rachh}

\author[4]{John C. Schotland}

\affil[1]{{\footnotesize Department of Statistics, University of Chicago, Chicago,
IL 60637, USA}}
\affil[2]{{\footnotesize Center for Computational Mathematics, Flatiron Institute, New York, NY 10010, USA}}
\affil[3]{{\footnotesize Center for Computational Quantum Physics, Flatiron Institute, New York, NY 10010, USA}}
\affil[4]{{\footnotesize Department of Mathematics, Yale University, New Haven, CT
06511, USA}}
\date{}

\begin{document}

\maketitle

\begin{abstract}
 In this paper we analyze the dynamics of
  single-excitation states, which model the scattering of a single
  photon from multiple two level atoms. For short times and weak
atom-field couplings we show that the atomic amplitudes are given by a
  sum of decaying exponentials, where the decay rates and Lamb shifts
  are given by the poles of a certain analytic function. This result is
  a refinement of the ``pole approximation'' appearing in the standard
  Wigner-Weisskopf analysis of spontaneous emission. On the other hand, at large times, the
  atomic field decays like $O(1/t^3)$ with a
  known constant expressed in terms of the coupling parameter and the
  resonant frequency of the atoms. Moreover, we show that for
  stronger coupling, the solutions also feature a collection of
  oscillatory exponentials which dominate the behavior at long times. Finally, we
  extend the analysis to the continuum limit in which atoms are
  distributed according to a given density. 
\end{abstract}

\section{Introduction}

Recent progress in experimental quantum optics has enabled the physical
construction of systems of ever-increasing complexity~\cite{Haroche_2006,Gardiner_2015,Liao_2016,Roy_2017,Kira_2011,Kimble_2008,Riedmatten_2008,Bloch_2012}. Of particular
interest is the scattering of one or two photons from a collection of
atoms. In this setting a central objective is to understand the time
evolution of the entanglement between atoms, mediated by the field.
Additionally, the ability to approximate the dynamics of these systems
numerically in an efficient and accurate manner is essential for
developing tools and theory for systems involving two or more entangled
photons. Questions of this nature will likely be at the heart of future
developments in a number of contexts, such as spectroscopy, imaging, and
communications.

We take as our starting point the model, proposed in \cite{kraisler21},
which involves the quantization of both the matter and the field. A
novel feature of this approach is that the electromagnetic field is
quantized in real space, putting the field and atomic degrees of freedom
on equal footing. In this model, if the matter consists of two-level
atoms and one makes the {\it rotating wave approximation}, then the
states involving one excitation (or one photon) decouple from those
which contain multiple photons or in which multiple atoms are
excited. These {\it single-excitation} states are the primary focus of
this paper. See \cite{kraisler2021one} for a discussion of the
two-photon problem.

In this paper we analyze the behavior of
single-excitation states with multiple atoms for intermediate and large
times. For the case of a single atom, similar analysis has been carried out 
for a variety of atom-field couplings by Knight and Milonni~\cite{knight}, Seke 
and Herfort (see \cite{seke88} and \cite{seke} for example), and Berman and Ford~\cite{berman2010}.


The structure of this paper is as follows. In Section 2 we review the
model for single-excitation systems with multiple atoms and derive an
integro-differential equation for the atomic amplitudes. Section 3
states the main result of this paper -- an asymptotic
expansion for the behavior of these systems at intermediate and large
times. A brief sketch of the proof is provided in Section 4, and the
full proof is developed in Section 5. In Section 6 we discuss the
continuum limit, in which the number of atoms is taken to infinity.

\section{Model}
We consider the following model for the interaction between a quantized
field and a system of $N$ two-level atoms located at $\bx_1,\dots,\bx_N
\in \bR^{3}.$ The atoms are taken to be stationary and sufficiently
well-separated so that their interactions can be neglected. We further
suppose that initially only one atom is in an excited state and that no
photons are present. Let $a_{j}(t)$ denote the probability amplitude for
the $j$th atom being in its excited state at time $t$, and $\psi(\bx,t)$
denote the wavefunction for the photon. In \cite{kraisler21}, it was
shown that within the rotating wave approximation, $a_1,\dots,a_N$ and
$\psi$ satisfy the following system of coupled equations
\begin{equation} \label{eq:sys}
  \begin{aligned}
i \frac{\partial}{\partial t} \psi(\bx,t) &= c \sqrt{-\Delta} \psi(\bx,t) + \sum_{j=1}^N a_j(t) \,\left(\Phi\star\delta(\cdot-\bx_j)\right)(\bx),\\
i \frac{{\rm d}}{{\rm d}t}a_j(t)&=\Omega a(t) +(\Phi\star \psi(\cdot,t))(\bx_j), \quad j=1,\dots,N,
  \end{aligned}
\end{equation}
together with suitable initial conditions, and boundary conditions at
infinity. Here $c$ denotes the speed of light, $\Omega>0$ is the
resonant frequency of the atoms, and $\hat{\Phi}(\bk),$ the Fourier
transform of $\Phi,$ is the coupling between the atom and the photon state
with wavenumber $\bk.$ In order for $\psi(\bx_j)$ to be
pointwise-defined we require $\hat{\Phi}$
to decay sufficiently rapidly at infinity. For concreteness, here we
treat the specific case in which $\Phi(\bx) = g
\exp(-|\bx|^2/(2R^2))/\sqrt{2\pi R^2}^3$ for some fixed positive
constants $g$ and $R,$ though the method we present generalizes in a
straightforward manner to a large class of couplings. With this choice,
the above equations become
\begin{align}
i \frac{\partial}{\partial t} \psi(\bx,t) &= c \sqrt{-\Delta} \psi(\bx,t) + \frac{g}{\sqrt{2\pi}^3R^3}\sum_{j=1}^N a_j(t) e^{-|\bx-\bx_j|^2/(2R^2)},\label{eqn:psi_a_psi}\\
i \frac{{\rm d}}{{\rm d}t}a_j(t)&=\Omega a(t) + \frac{g}{\sqrt{2\pi}^3R^3} \int e^{-|\bx-\bx_j|^2/(2R^2)} \psi(\bx,t)\,{\rm d}\bx \quad j=1,\dots,N\label{eqn:psi_a_a}.
\end{align}
Equation (\ref{eqn:psi_a_psi}) can be used to write $\psi$ in terms of $a,$ which yields
$$\psi(\bx,t) = -\frac{ig}{(2\pi)^3} \sum_{j=1}^N \int_0^t a_j(\tau) \int_{\mathbb{R}^3} e^{-ic k(t-\tau)-k^2R^2/2}e^{i\bk\cdot(\bx-\bx_j)}{\rm d} \bk \,{\rm d}\tau.$$
Here we adopt the standard convention of using regular font to denote the norm of the corresponding vector quantity (e.g. $k=|\bk|$). We reserve the use of the vector symbol `$\,\,\vec{\cdot}\,\,$' for vectors in $\mathbb{C}^N,$ with elements indexed by atom number.

After substituting this expression for $\psi$ into (\ref{eqn:psi_a_a}) we obtain the following system of coupled integro-differential equations for the atomic amplitudes $a_j,$
\begin{equation}
  \begin{aligned}
i\frac{{\rm d}}{{\rm d}t} a_j(t)=&\Omega a(t) -\frac{g}{\sqrt{2\pi}^3R^3}  \int e^{-|\bx-\bx_j|^2/(2R^2)}    \\
& \quad\times   \frac{ig}{(2\pi)^3} \sum_{\ell=1}^N \int_0^t a_\ell(\tau) \int_{\mathbb{R}^3} e^{-ick(t-\tau)-k^2R^2/2}e^{i\bk\cdot(\bx-\bx_\ell)}{\rm d} \bk \,{\rm d}\tau {\rm d} \bx.
  \end{aligned}
\end{equation}
After performing the integral in $\bx$, and dividing by $i,$ the above equations simplify to
\begin{align}
\label{eq:asystem}
\frac{{\rm d}}{{\rm d}t} a_j(t)=&-i\Omega a(t) -   \frac{g^2}{(2\pi)^3} \sum_{\ell=1}^N \int_0^t a_\ell(\tau) \int_{\mathbb{R}^3} e^{-ick(t-\tau)-k^2R^2}e^{i\bk\cdot(\bx_j-\bx_\ell)}{\rm d} \bk \,{\rm d}\tau .
\end{align}
Here, for ease of exposition, we have assumed that all atoms have identical resonant frequencies. Our analysis may be extended in a straightforward manner to allow for variations in the atomic resonant frequencies. 

In order to simplify the equations, we let $\beta_{j}(t) = e^{i \Omega t} a_{j} (t)$, and observe that
\begin{align}
\frac{{\rm d}}{{\rm d}t}\beta_j(t) = -\frac{g^2}{(2\pi)^3} \sum_\ell \int_0^t  \,\beta_\ell(\tau)e^{i\Omega (t-\tau)} \int_{\bR^{3}}  e^{-ic|\bk|(t-\tau)- |\bk|^2 R^2}e^{i\bk\cdot(\bx_j-\bx_\ell)}{\rm d}\bk\,{\rm d}\tau.
\end{align}
Let $r_{j,\ell} = |\bx_j-\bx_\ell|$ and let $f: \bR^{2} \to \bR$ be given by
\begin{align}
f(k,r) = \int_{S^{2}} {\rm d}\hat{\theta} \, e^{ik r\, \cos(\hat{\boldsymbol \theta} \cdot \hat{\bx})} = \frac{4\pi \sin{(kr)}}{kr} \, .
\end{align}
Then
\begin{align}\label{eqn:beta_main}
\frac{{\rm d}}{{\rm d}t}\beta_j(t) = -\frac{g^2}{(2\pi)^3} \sum_\ell \int_0^t  \,\beta_\ell(\tau)e^{i\Omega (t-\tau)} \int_0^\infty \,k^2 e^{-ick(t-\tau)-k^2 R^2} f(k,r_{j,l})\,{\rm d}k\, {\rm d}\tau.
\end{align}
It is this system of integro-differential equations which we take as the starting point of our analysis.

\section{Statement of the main result}
In this section we provide a brief description of the main results.
\begin{theorem}
Let $\vec{\beta}$ be the solution to (\ref{eqn:beta_main}) and $1 \le N <\infty.$ Then
\begin{equation} \label{eqn:main_result}
  \begin{multlined}
\vec{\beta}(t) = \sum_{j=1}^{n_p} R_j e^{i(p_j+\Omega)t}
    \vec{\beta}(0) + \sum_{j=1}^{n_s} S_j(t) e^{i(-z_j+\Omega)t}
    \vec{\beta}(0) \\ - \frac{g^2e^{i \Omega t}}{2\pi^2 ic^2 \Omega^2 t^3} \vec{\beta}(0) + O\left( \frac{g^4}{t^3}\right)\vec{\beta}(0) +o\left( \frac{g^2}{t^3}\right) \vec{\beta}(0)+O\left(\frac{g^2 \epsilon^2 e^{-\epsilon t/2} }{\Omega^2ct}\right)\vec{\beta}(0),
  \end{multlined}
\end{equation}
as $t \to \infty$ with $g$ held fixed and sufficiently small. Here
\begin{enumerate}
\item $\epsilon$ is a positive real number independent of $t$ and $g$
\item The $p_j$ are the $n_p \le N$ positive real numbers satisfying
$${\rm det} \left[(p_j+\Omega)I-\frac{g^2}{(2\pi)^3}\int_0^\infty \frac{(ck+p_j) e^{-k^2R^2}}{(ck+p_j)^2} k^2 f(k,r_{j,\ell}) {\rm d}k \right] = 0,$$
 and the matrices $R_j$ are of the form $c_j \vec{v}_j \vec{v}_j^*$ where $\vec{v}_j$ is the unit vector such that
$$\left[(p_j+\Omega)I -\frac{g^2}{(2\pi)^3}\int_0^\infty  \frac{(ck+p_j) e^{-k^2R^2}}{(ck+p_j)^2} k^2 f(k,r_{j,\ell}){\rm d}k\right] v_j= 0,$$
    and $c_j$ is a scalar depending only on $j,g,R,c,$ and $\Omega.$
\item The $z_j$ are the $n_s\le N$ distinct poles of the entries of the $N\times N$ matrix
$$H^-(y) = \left( y- \Omega + \frac{g^2}{(2\pi)^3c} A^-(y) \right)^{-1},$$
in the fourth quadrant, where $A^-$ is a matrix-valued function defined in~\cref{eq:F_def}, which is analytic everywhere in the complex plane except for a branch cut along the negative real axis. The matrices $e^{-iz_jt} S_j(t)$, $j=1,\dots,n_s$, are the residues of $e^{-iyt} H^-(y)$ corresponding to the poles $z_1,\dots,z_{n_s}.$ Moreover, if the poles are simple then $n_s = N$ and the $S_j$ are constant.
\end{enumerate}
\end{theorem}
We conclude this section with several remarks pertaining to extensions of this result, and connections to the literature.
\begin{rem}
  The second term in \eqref{eqn:main_result} is an improvement on
  the ``pole approximation'', which is obtained by approximating the $z_j$ using a perturbative expansion of $A^-(y)$ about $y= \Omega.$ In particular, for small $g,$ the $z_j$ are well-approximated by the poles of the matrix
$$\left(y-\Omega + \frac{g^2}{(2\pi)^3c}A^-(\Omega)\right)^{-1}. $$
Similarly, the $S_j$ are well-approximated by the corresponding residues. When $g$ is sufficiently small, standard perturbation theory arguments show that this simplification provides a reasonable approximation to
  $z_j$ and $S_j$.
\end{rem}
\begin{rem}
In~\cref{eqn:main_result} the coefficient in front of $g^2/t^3$ can be improved. We refer the reader to~\cref{eqn:alpha1} and the surrounding discussion for more details.
\end{rem}
\begin{rem}
The single atom case was considered in \cite{berman2010} for a variety of coupling functions $\Phi.$ This paper consists of both an extension of that result to multiple atoms, as well as a different derivation of the pole approximation which yields a single expression valid both in the algebraic and exponential decay regimes.
\end{rem}

\begin{rem}
\label{rem:flopcount}
For small $g$, the poles $z_{j}$, $j=1,2,\ldots N$ the numbers $p_{j}$, $j=1,2,\ldots n_{p} \leq N$, and the corresponding residues $S_{j}$ and $R_{j}$ can be computed using a root finding algorithm such as Newton's method or secant method. For each pole $z_{j}$, the method typically converges in $O(1)$ iterations with each iteration requiring the computation of the inverse of an $N \times N$ matrix. Thus, all of the quantities $z_{j},p_{j},S_{j}$, and $R_{j}$ can be computed in $O(N^4)$ operations. Finally, given these quantities, the approximation of $\beta(t)$ through~\cref{eqn:main_result} can be computed in $O(N)$ operations for any time $t$.
\end{rem}

%

\section{Idea of the proof}
\begin{figure}[h]
\centering
\includegraphics[width=0.35\textwidth]{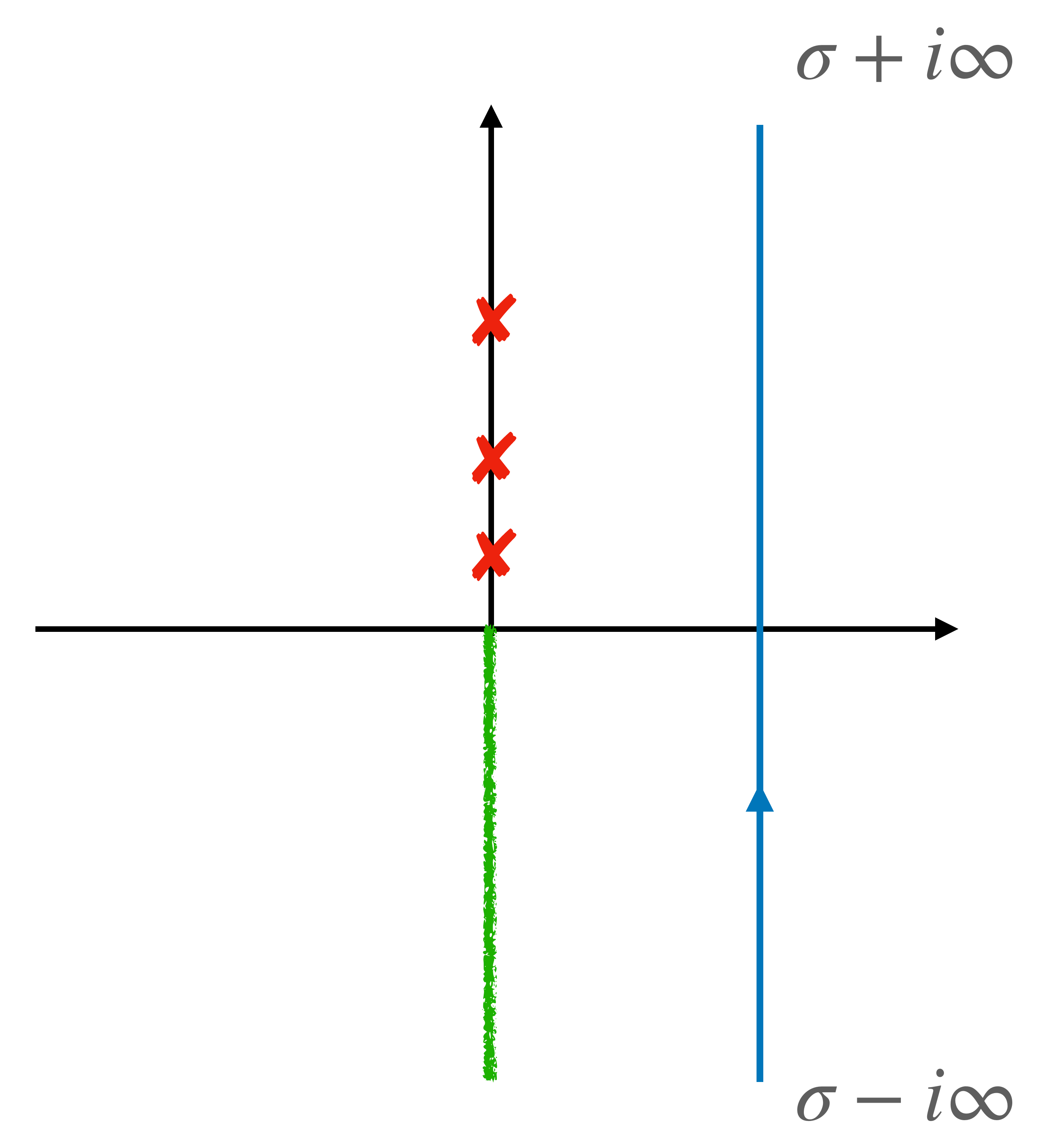}
  \caption{The contour for the inverse Laplace transform (blue), the
  poles of the integrand on the positive imaginary axis (red), and the
  branch cut of the integrand, lying along the negative imaginary axis
  (green).}\label{fig:lap_inv}
\end{figure}

In this section we give a brief description of the idea of the proof. As
in \cite{berman2010}, we solve the integro-differential equation
(\ref{eqn:beta_main}) by Laplace transforms. Upon taking a Laplace
transform in $t$ we arrive at a system of linear equations for the
Laplace transforms of $\beta_1,\dots,\beta_N$ which we denote by
$B_1,\dots,B_N$ respectively. The original variables
$\beta_1,\dots,\beta_N$ can then be obtained by integrating
$B_1(s),\dots,B_N(s)$ along the contour $\sigma +i \mathbb{R}$, for
$\sigma$ sufficiently large that all the poles of $B_1,\dots,B_N$ lie to the left of the contour; see Figure \ref{fig:lap_inv}.

\begin{figure}[h!]
\centering
\includegraphics[width=0.35\textwidth]{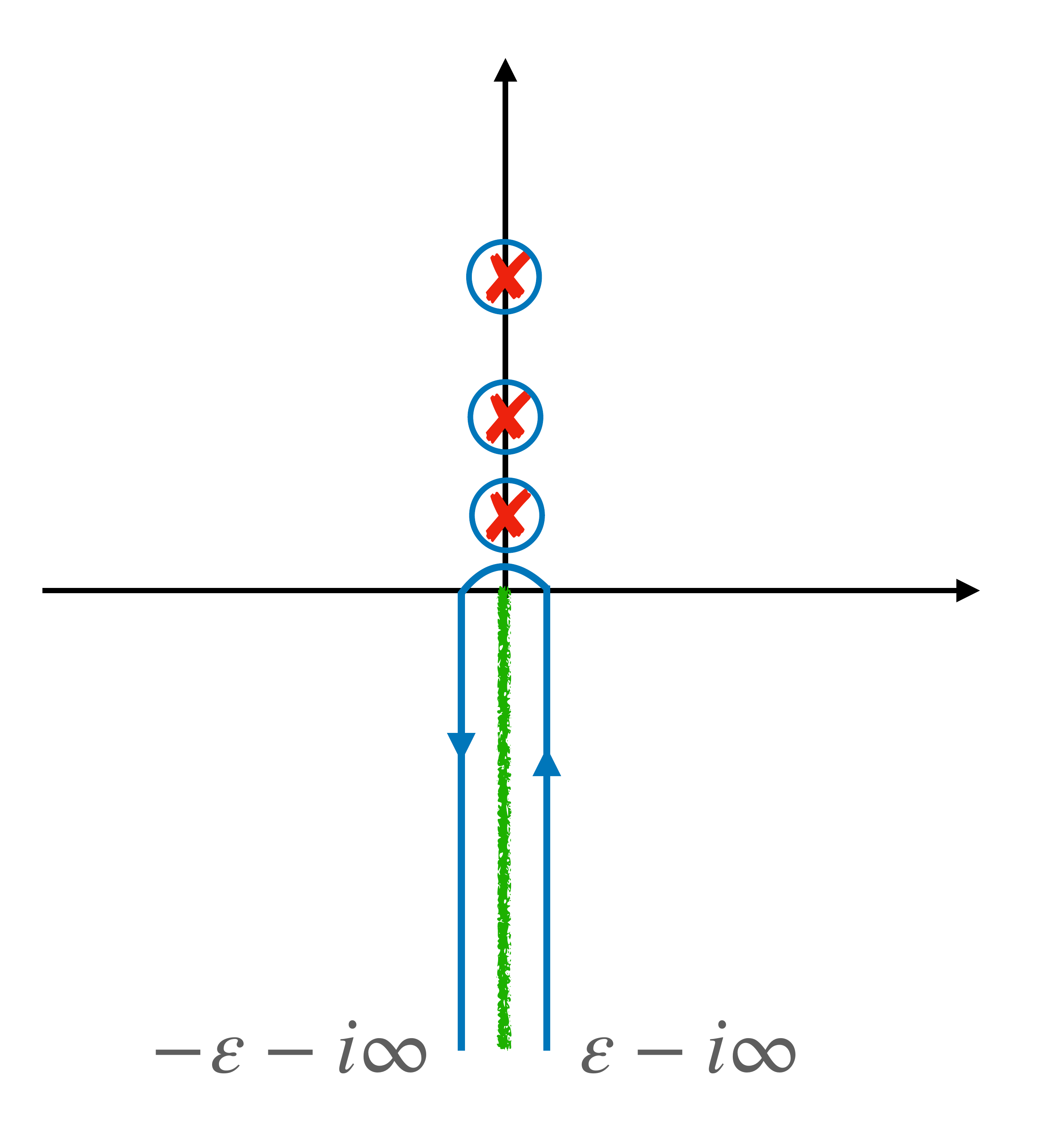}
  \caption{The deformed contour for the inverse Laplace transform
  (blue), the poles of the integrand on the positive imaginary axis
  (red), and the branch cut of the integrand, lying along the negative
  imaginary axis (green).}\label{fig:lap_inv2}
\end{figure}

Next, we show that there are at most $N$ poles of the integrand, all of which lie on the positive imaginary axis. On the negative imaginary axis there is a branch cut (see Figure \ref{fig:lap_inv}). Then we deform the contour to the one shown in Figure \ref{fig:lap_inv2}. The contributions of the isolated poles on the positive imaginary axis correspond to states that oscillate but do not decay. For sufficiently small $g,$ and with all other parameters held fixed, there are no such poles.

\begin{figure}[h!]
\centering
\includegraphics[width=0.5\textwidth]{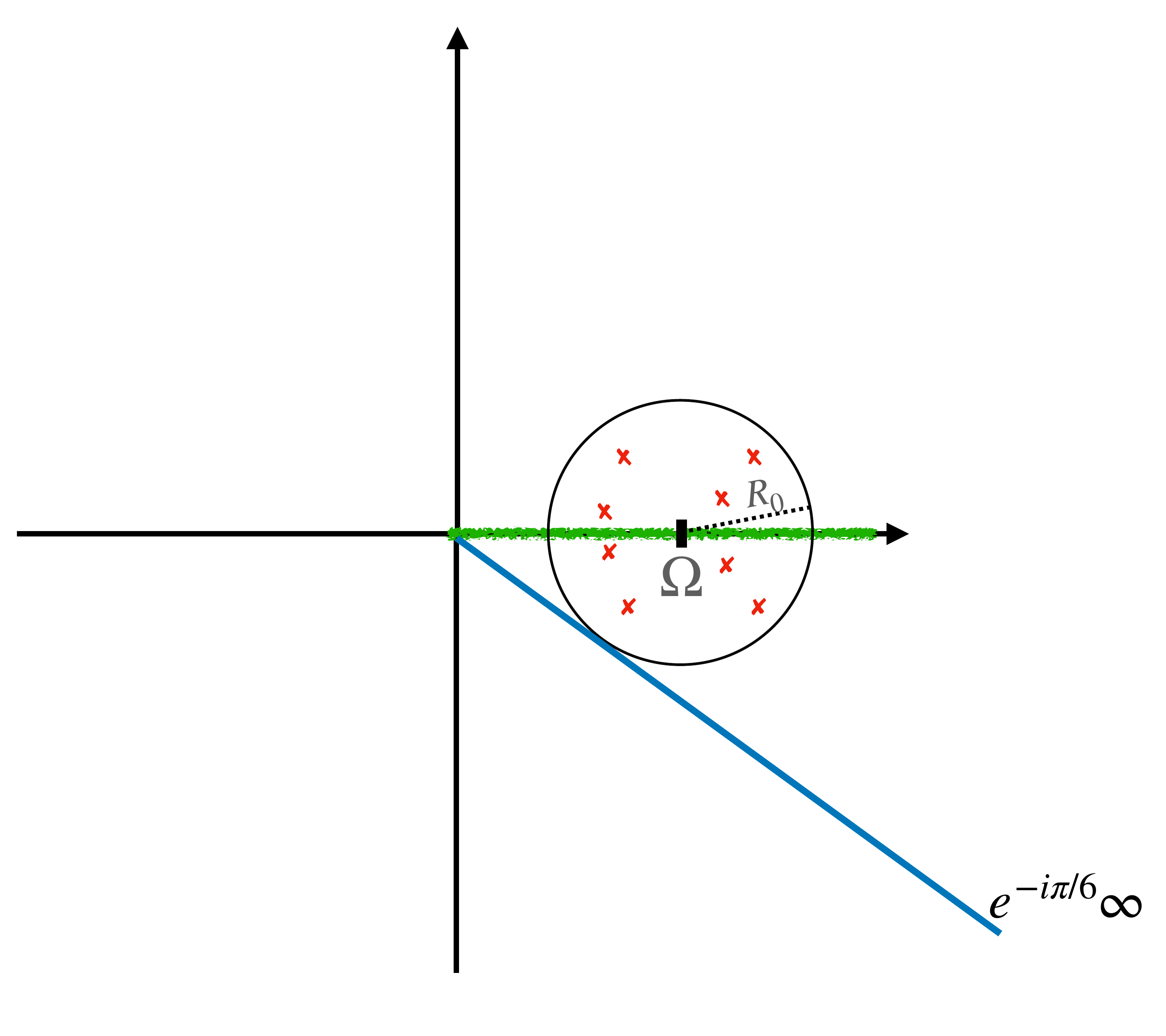}
\caption{The contour for the integral along the branch cut after a
  change of variables (green), the poles of the new integrand (red), and
  the final deformed contour along which the asymptotic expansion is
  computed (blue).}\label{fig:lap_inv3}
\end{figure}

Next we proceed by writing the integrals on either side of the branch cut as a single integral, and make a change of variables so that the domain of integration is the positive real axis, as in Figure \ref{fig:lap_inv3}. We compute the asymptotic behavior of this integral by deforming the contour down to the ray $e^{-i\pi/6}\mathbb{R}_+.$ Along this ray, the integrand decays exponentially in $tk$ where $t$ is the time and $k$ is the variable of integration. An asymptotic expansion can then be obtained by repeated integration by parts.

We must include the contribution of any poles
lying in the region between the original and final contours. We show that for $g$ sufficiently small, there exists an $R_0$ such that the integrand has $2N$ poles in the disk of radius $R_0$ centered at $\Omega.$ Half of these poles lie in the upper half plane and the other half lie in the lower half plane. Note that these poles are not poles of the original integrand for the inverse Laplace transform. They correspond to poles of the integrand along the branch cut after the contributions from both sides have been combined.

\section{Analytical apparatus}
In our analysis we follow a similar approach to that described in \cite{berman2010}. We begin by taking the Laplace transform of (\ref{eqn:beta_main}). Setting $B_{j}(s) = \cL[\beta_{j}](s) = \int_{0}^{\infty} e^{-st} \beta_{j}(t) dt$,  we obtain
\begin{align}
s B_j(s)-\beta_j(0) = - \frac{g^2}{(2\pi)^3} \sum_\ell B_\ell(s) \Gamma_{j,\ell}(s-i\Omega),
\end{align}
where
\begin{equation}
  \begin{aligned}
\Gamma_{j,\ell}(s) &= \int_0^\infty \int_0^{\infty}  \,e^{-ickt-st - k^2 R^2} k^2 f(k,r_{j,\ell}){\rm d}k \,{\rm d} t\\
&=\frac{1}{ic} \int_0^{\infty}  \frac{e^{-k^2R^2}}{\frac{s}{ic}+k} k^2 f(k,r_{j,\ell})\,{\rm d}k
  \end{aligned}
\end{equation}

Next, we invert the Laplace transform to obtain the solution $\beta_{j}(t)$ in terms of the functions $\Gamma_{j,\ell},$ which produces  the following formulae for $\beta_1,\dots,\beta_N,$
\begin{equation}
  \begin{aligned}
\vec{\beta}(t) &= \frac{1}{2\pi i} \int_{\sigma -i\infty}^{\sigma+i\infty} e^{st}\left(s I +\frac{g^2}{(2\pi)^3} \Gamma(s-i\Omega) \right)^{-1}\,{\rm d}s \,\vec{\beta}(0)\\
&=\frac{e^{i\Omega t}}{2\pi i} \int_{\sigma -i\infty}^{\sigma+i\infty} e^{st}\left(s I +i\Omega I+\frac{g^2}{(2\pi)^3} \Gamma(s) \right)^{-1}\,{\rm d}s\,\vec{\beta}(0) \label{eq:lapinv}, 
\end{aligned}
\end{equation}
where $\vec{\beta}(t) = [\beta_{1}(t),\beta_{2}(t) \ldots
\beta_{N}(t)]^{T}$, $\Gamma$ is an $N \times N$ matrix whose entries are $\Gamma_{j,\ell},$ and $\sigma$ is a sufficiently large positive real number. 

\begin{lemma}
\label{lem:Gammadef}
For each $j,\ell$, $\Gamma_{j,\ell}(s)$ is analytic for
  $s \in \mathbb{C} \backslash L$, where $L$ is the negative imaginary
  axis. There, $\Gamma_{j,\ell}(s)$ has a branch cut. Furthermore, if $0<y$,
\begin{equation}\label{eq:F_def}
  \begin{aligned}
\lim_{\varepsilon \to 0^{+}} ic\Gamma_{j,\ell} (-iy \pm \varepsilon)
    &= \left(  \pm e^{-y^2R^2/c^2}y^2 f(y/c, r_{j,\ell})  i\pi +f(y/c,
  r_{j,\ell}) R^2\Theta(yR/c)+ G_{j,\ell}(y)\right) \\
    &\equiv A^{\mp}(y),
  \end{aligned}
\end{equation}
where 
$$\Theta\left(y\right) = {\rm p.v.} \int_0^\infty k^2 e^{-k^2} (k-y)^{-1}{\rm d}k= y^2\log(y) + O(y^3) \, ,\quad \text{as } y\to 0\,,$$
and 
$$G_{j,\ell}(y):= \int_0^\infty k^2 e^{-k^2R^2}
  \frac{[f(k,r_{j,\ell})-f(y/c,r_{j,\ell})]}{k-\frac{y}{c}}\,{\rm d}k$$
  are analytic in the right-half plane  and are real on the real axis.
Moreover $A^{+}(y) =
  \overline{A^{-}(\bar{y})}$ for all $y\in \mathbb{C}\setminus \mathbb{R}_-.$
\end{lemma}
The following lemma provides bounds on $A^{\pm}$ for a certain region in the complex plane. Its proof follows in a straightforward manner from the definition of $A^{\pm}$ and is omitted.
\begin{lemma}\label{lem:omegaf}
Consider the region in the complex plane $\Omega_A$ defined by
\begin{align}
\Omega_A:= \{ z = r e^{i\theta}\,| \, r>0, -\pi/6 \le\theta\le\pi/6\}.
\end{align}
Then there exists a positive constant $C_F$ depending on $x_1,\dots,x_N,$ $c,$ $R,$ and $\Omega$ such that 
\begin{align}
\sup_{y \in \Omega_A,\,j,\ell=1,\dots,N} |A^{\pm}_{j,\ell}(y)| <C_A.
\end{align}
\end{lemma}

We next prove an elementary result regarding the positive definiteness of a family of matrices related to integrals of $f(k,r_{j,\ell})$.
\begin{lemma}
\label{lem:posdef}
Suppose that the points $\bx_{j} \in \mathbb{R}^{3}$ are distinct and that $h(k):[0,\infty) \to \mathbb{R}$ is smooth, bounded and positive. Then the matrix $E$ whose entries $E_{j,\ell}$ are
\begin{equation}
E_{j,\ell} = \int_{0}^{\infty} {\rm d}k \, h(k) f(k,r_{j,\ell}) \, ,
\end{equation}
is positive definite.
\end{lemma}
\begin{proof}
Let $v \in \mathbb{C}^{N}$, and consider the quadratic form $v^{\star} E v$ \,, 
\begin{equation}
\begin{aligned}
v^{\star}Ev &= \sum_{j=1}^{N} \sum_{\ell=1}^{N} \int_{0}^{\infty} {\rm d}k \, h(k) f(k,r_{j,\ell}) v_{j} \overline{v}_{\ell}  \, ,\\
&= \sum_{j=1}^{N} \sum_{\ell=1}^{N} \int_{\bR^{3}} {\rm d} k\, h(|k|) e^{i k \cdot( x_{j} - x_{\ell})} v_{j} \overline{v}_{\ell} \, \\
&= \int_{\bR^{3}} {\rm d} k \, h(|k|) \left|\sum_{j=1}^{N} e^{ik \cdot x_{j}} v_{j} \right|^2  \, ,
\end{aligned}
\end{equation}
where in going from the first to the second inequality we have used the definition of $f.$ The above integral is clearly non-negative since $h(k)>0$ on the domain of definition. Moreover if $v^{\star} Ev = 0$, then
$\sum_{j=1}^{N} e^{ik\cdot x_{j}} v_{j} =0$ for all $k \in \bR^{3}$.  However, since the $x_{j}$ are distinct, the functions $e^{ik \cdot x_{j}}$ are linearly independent functions of $k$, and hence the above identity holds for all $k \in \bR^{3}$ if and only if $v = 0$.
\end{proof}

\begin{remark}
An almost identical argument can be used to show that $f(k_0,r_{j,\ell})$ is positive definite for any $k_0 >0.$
\end{remark}

 Since $(s + i\Omega)I + \frac{g^2}{(2\pi)^3} \Gamma(s)$ is an
analytic matrix-valued function on $\mathbb{C} \backslash L$, its inverse is a meromorphic
matrix-valued function on the same domain. In order to deform the contour of integration in~\cref{eq:lapinv} to an
integral along the branch cut (which lies on the negative imaginary
axis), we must determine the poles and corresponding residues of $\left((s +
i\Omega)I + \frac{g^2}{(2\pi)^3} \Gamma(s)\right)^{-1}$ in the left half
plane, including the positive imaginary axis. A characterization of the number and locations of these poles is given by the following lemma.

\begin{lemma}\label{lem:gam_mero}
The matrix-valued function $\left((s +
i\Omega)I + \frac{g^2}{(2\pi)^3} \Gamma(s)\right)^{-1}: \mathbb{C} \setminus L \to
  \mathbb{C}^{N\times N}$ is meromorphic on its domain of definition,
  and its poles lie on the positive imaginary axis.
\end{lemma}

\begin{proof}
We begin by recalling that
\begin{align*}
\Gamma_{j,\ell}(s) =\frac{1}{ic} \int_0^{\infty} {\rm d}k \frac{e^{-k^2R^2}}{\frac{s}{ic}+k} k^2 f(k,r_{j,\ell}).
\end{align*}
For ease of exposition we define $f_k$ to be the $N\times N$ matrix with
  the $j,\ell$ entry given by $f(k,r_{j,\ell}).$

Now, we note that a point $s\in \mathbb{C} \setminus L$ is a pole of
  $(sI + i \Omega I + \frac{g^2}{(2\pi)^3}\Gamma(s))^{-1}$ if and only if  
$sI + i \Omega I + \frac{g^2}{(2\pi)^3} \Gamma(s)$  is not invertible. 
Substituting $s = x + iy$ we see that
\begin{equation}
\begin{aligned}
(x+iy)I +i\Omega I +\frac{g^2}{(2\pi)^3}& \Gamma(x+iy) = xI +\frac{g^2}{(2\pi)^3} \int_0^{\infty}k^2\frac{x e^{-k^2 R^2}}{x^2+(ck+y)^2} f_k\,{\rm d}k\\
&+ i \left[(y+\Omega)I -\frac{g^2}{(2\pi)^3} \int_0^{\infty}k^2 \frac{(ck+y)e^{-k^2 R^2}}{x^2+(ck+y)^2} f_k\,{\rm d}k\right]\\
&= A_R +iA_I,
\end{aligned}
\end{equation}
where $A_R$ and $A_I$ are both real, symmetric matrices.

Hence, if $s=x+iy$ is a pole then there exists $\vec{v} \in \mathbb{C}^{N}$, with $\|\vec{v}\|=1$, such that 
\begin{align}
\vec{v}^{\star} A_{R} \vec{v} = \vec{v}^{\star} A_{I}\vec{v}  = 0 \, .
\end{align}
Looking at the real part $\vec{v}^{\star} A_R\vec{v}$, we get
\begin{align}
x\left( 1 + \frac{g^2}{(2\pi)^3} \int_0^{\infty}k^2 \frac{e^{-k^2 R^2}}{x^2+(ck+y)^2}  \vec{v}^* f_k \vec{v}\,{\rm d}k \right) = 0.
\end{align}
Next, we observe that  $\int_0^{\infty}{\rm d}k\, \frac{e^{-k^2
  R^2}}{x^2+(ck+y)^2}  \vec{v}^* f_k \vec{v} \geq 0$ and thus the above equation holds if and only if $x=0$.
\end{proof}

In light of the previous lemma, in order to determine the poles of $sI + i \Omega I + \frac{g^2}{(2\pi)^3} \Gamma(s)$ we need only consider points $s = iy,$ $y>0,$ on the positive imaginary axis. Plugging in $s=iy,$ we see that
$$sI + i \Omega I + \frac{g^2}{(2\pi)^3} \Gamma(s) =i ((y + \Omega)I - E(y)),$$
where
\begin{equation}
E_{j,\ell}(y) = i \frac{g^2}{(2\pi)^3}\Gamma_{j,\ell}(iy) = \frac{g^2}{(2\pi)^3}\int_{0}^{\infty}  \,k^2 \frac{f(k,r_{j,\ell}) e^{-k^2 R^2}}{y+ck} {\rm d}k\, , \quad y > 0 \, .
\end{equation}
Since, the matrix $E(y)$ is symmetric, it is always diagonalizable and its
eigenvectors are orthonormal. We let $\lambda_{j}(y)$ denote its
eigenvalues and $v_{j}(y)$ the corresponding eigenvectors.

\begin{lemma}
Suppose $y_{0}$ satisfies $\lambda_{j}(y_{0}) = y_{0} + \Omega$, then $((y + \Omega) I - E(y))^{-1}$ has a pole at $y=y_{0}$ and the corresponding residue is given by 
\begin{equation}
R(y_{0}) = \frac{ \vec{v}_{j}(y_{0}) \vec{v}_{j}^{\star}(y_{0})}{1-\lambda_{j}'(y_{0})} ,
\end{equation}
where $\vec{v}_j$ is the corresponding unit eigenvector. Moreover, for each $j=1,\dots,N$ there exists at most one zero of $\lambda_{j}(y) = y+\Omega$ on $(0,\infty)$. 
\end{lemma}
\begin{proof}
Letting $V= (\vec{v}_1,\dots,\vec{v}_N)$ denote the matrix of eigenvectors, we observe that
\begin{equation}
y + \Omega I - E(y) = V^{\star} \begin{bmatrix} 
y + \Omega - \lambda_{1}(y)  & & & \\
& y + \Omega - \lambda_{2}(y) & & \\
& & \ddots & \\
& & & y + \Omega - \lambda_{N}(y) 
\end{bmatrix} V \,.
\end{equation}
Therefore, $((y + \Omega)I - E(y))^{-1}$ has a pole whenever $y + \Omega - \lambda_{j}(y) = 0$ for some $j.$
Moreover, we observe that 
\begin{equation}
\lambda_{j}(y) = \vec{v}_{j}^{\star}(y) E(y) \vec{v}_{j}(y) \, .
\end{equation} 
A simple calculation shows that
\begin{equation}
\label{eq:lamb-der-def}
\lambda_{j}'(y) = \lambda_j(y) \,\left[(\vec{v}_j^\star(y))' \vec{v}_j(y) +\vec{v}_j^\star(y)\vec{v}_j'(y)\right] -\frac{g^2}{(2\pi)^3}\int_0^{\infty} k^2 \frac{e^{-k^2 R^2}}{(ck+y)^2} \vec{v}_j^\star(y) f_k \vec{v}_j(y)\,{\rm d}k.
\end{equation}
The first term is always zero since $\|\vec{v}_{j}(y)\|=1$ and the second term is negative, since $\langle \vec{v}_j f_k \vec{v}_j\rangle >0$ if $\vec{v}_j \neq 0.$ Thus $\lambda_{j}'(y) < 0$ for all $y$.
Therefore the pole of the matrix $(y+\Omega)I - E(y)$ at $y=y_{0}$ associated with $y_{0} + \Omega = \lambda_{j}(y_{0})$ is a simple pole whose residue is given by
\begin{equation}
R(y_{0}) = \lim_{y \to y_{0}} (y-y_{0}) ((y-\Omega)I - E(y))^{-1} = \left( \frac{ \vec{v}_{j}(y_{0}) \vec{v}_{j}^{\star}(y_{0})}{1-\lambda_{j}'(y_{0})} \right)\, .
\end{equation}

Finally, to see that $\lambda_{j}(y) = y+\Omega$ on $(0,\infty)$ has at most one zero, we observe that by~\cref{eq:lamb-der-def},  $\lambda_{j}(y)$ is a monotonically decreasing function of $y.$ This in turn implies that $y + \Omega - \lambda_{j}(y)$ is monotonically increasing from which the result follows.
\end{proof}

The following result provides a necessary condition for the existence of poles.

\begin{lemma}
There exists a constant $C$  such that $(s+i\Omega I + \frac{g^2}{(2\pi)^3} \Gamma(s))^{-1}$ has no poles whenever
$$N\frac{g^2}{c \Omega R^2} <C.$$
\end{lemma}
\begin{proof}
The proof follows immediately by observing that
$$\left|\int_{0}^{\infty} \,k^2 \frac{f(k,r_{j,\ell}) e^{-k^2 R^2}}{y+ck}{\rm d}k\right| \le 4\pi \int_{0}^{\infty} \, k^2 \frac{e^{-k^2 R^2}}{y+ck}\,{\rm d}k \le \frac{2\pi}{cR^2}.  $$
and applying the Gershgorin circle theorem.
\end{proof}
\begin{remark}
Note that the above theorem does
  not exclude the possibility that the eigenvalue corresponding to a
  particular pole $y_0$ is degenerate, in which case $\lambda_j(y_0) =
  y_0 + \Omega$ for more than one values of $j$. However, in this case, there is still a
  set of orthonormal eigenvectors, and the expression for the residues
  is unchanged.
\end{remark}

Combining all the results above, we can now deform the contour of integration for computing the solution $\vec{\beta}(t)$ from the vertical line $\sigma + i\mathbb{R}$ to the Bromwich contour to obtain the following result. Its proof is a straightforward application of the preceding results and Cauchy's integral theorem and is omitted.
\begin{theorem}
Suppose that $p_{j},$ $j=1,\dots,n_p$ are the poles of the matrix $(y + \Omega - E(y))^{-1}$ and $R_j$ are the corresponding residues. Then all such poles are positive and real, and $n_p$ is at most $N.$ Furthermore,
\begin{equation}
\begin{aligned}
\label{eq:betafin}
\vec{\beta}(t) = \frac{e^{i \Omega t}}{2\pi i} &\int_{0}^{\infty} e^{-iyt} \left(  \left( y - \Omega + \frac{g^2}{(2\pi)^3 c} A^{+}(y) \right)^{-1} - \left( y - \Omega + \frac{g^2}{(2\pi)^3 c} A^{-}(y) \right)^{-1} \right) \vec{\beta}(0) \, {\rm d}y \\
& +\sum_{j=1}^{n_{p}} R_j \vec{\beta}(0) e^{i(p_{j} + \Omega) t} ,\,  
\end{aligned}
\end{equation}
where $A^\pm$ are the matrices defined in (\ref{eq:F_def}).
\end{theorem}

We observe that the functions
\begin{equation}
H^{\pm}(y) = \left( y- \Omega + \frac{g^2}{(2\pi)^3 c} A^{\pm}(y) \right)^{-1} \, ,
\end{equation}
are analytic everywhere except the branch cut on the negative real axis. Thus we can deform the contour of integration in~\cref{eq:betafin} to a ray $\theta=-\pi/6$ in the fourth quadrant. In so doing we pick up a contribution from the poles of $H^{\pm}(y)$ with $-\pi/6 < {\rm arg}\, y < 0$. Specifically, we see that
$$\int_0^{\infty} e^{-iyt}(H^+(y)- H^-(y)) \, {\rm d}y = \int_0^{\infty \,e^{-i \pi/6}} e^{-iyt}(H^+(y)- H^-(y)) \, {\rm d}y + 2\pi i \sum_{j} e^{-i z_j t} S_j(t), $$
where $z_j,$ $j=1,2,\dots$ are the poles of $(H^+(y)- H^-(y))$ lying between the positive real axis and the ray $e^{-i\pi/6} [0,\infty).$ The $S_j(t)$ are the corresponding residues, defined via the formula
$$S_j(t) = \frac{e^{iz_jt}}{2\pi i}\lim_{\epsilon \to 0^+} \int_{|\xi-z_j|=\epsilon} e^{-i \xi t} (H^+(\xi)- H^-(\xi))\,{\rm d}\xi.$$ We note that the existence of the limits in the previous expression are guaranteed by the meromorphicity of $H^\pm.$

The following lemma provides bounds on the location and number of poles of $H^\pm.$

\begin{lemma}
Let $\Omega_A$ be the region defined in Lemma \ref{lem:omegaf}, and
  consider the matrix-valued function
\begin{align}
( y -  \Omega) I + \frac{g^2}{(2\pi)^3 c} A^{+}(y).
\end{align}
There is a positive constant $G_0$ (depending only on
  $\Omega,$ $c,$ and $\bx_1,\dots,\bx_N$) such that for all $g < G_0$ there exists an $R_0<\Omega
  \,\sin \frac{\pi}{8}$ and exactly $N$ roots $y_1,\dots,y_N \in D(\Omega,R_0)
  \cap \overline{\Omega}_A \cap \mathbb{C}^+$  counting multiplicity
  such that
\begin{align}
\det\left[ (y -  \Omega)I + \frac{g^2}{(2\pi)^3 c} A^{+}(y) \right]  = 0. 
\end{align}
\end{lemma}
\begin{proof}
We begin by observing that for $g>0,$ there are no roots on the positive real axis. Indeed, assume the contrary; namely, suppose there existed a $y_0 \in \mathbb{R}^+$ such that
$$(y_0-\Omega)I + \frac{g^2}{(2\pi)^3c} A^+(y_0)$$
  had a null-vector $\vec{v}_0.$ From \eqref{eq:F_def}, we see that
  $A^{+}(y) = A_R(y)+iA_I(y)$, where $A_R$ and $A_I$ are symmetric matrices and $A_I$ is negative definite. It follows that
\begin{align*}
  0&=\vec{v}_0^* \left[(y_0-\Omega)I + \frac{g^2}{(2\pi)^3c}
  A^+(y_0)\right]\vec{v}_0 \\
  &= \vec{v}_0^* \left[(y_0-\Omega)I + \frac{g^2}{(2\pi)^3c} A_R(y_0)\right]\vec{v}_0 +i\frac{g^2}{(2\pi)^3c} \vec{v}_0^* A_I(y_0) \vec{v}_0.
\end{align*}
Since $A_I$ is negative definite the imaginary part cannot vanish, which is a contradiction. Thus, for $g>0$ there are no roots on the positive real axis.

Next, we let $C_A$ denote the constant defined in Lemma \ref{lem:omegaf} and choose $R_0 = \Omega \sin\frac{\pi}{8}.$ If $G_0$ is chosen such that $\frac{2N G_0^2 C_A}{(2\pi)^3 c} < R_0$ it follows from the Gershgorin circle theorem that
\begin{align}
(y-\Omega)+\frac{g^2}{(2\pi)^3c}A^{+}(y)
\end{align}
is nonsingular for all $y \in D^c(\Omega,R_0)\cap \Omega_A.$ In particular, for all vectors $v \in \mathbb{C}^n,$
\begin{align}
\label{eq:matinvbound}
\left|\vec{v}^* \left[ (y-\Omega)I + \frac{g^2}{(2\pi)^3c} A^{+}(y) \right] \vec{v}\right|  \ge \frac{R_0}{2},
\end{align}
for $y \in \Omega_A,$ $|y-\Omega| \ge R_0.$

Now, we observe that the roots of 
\begin{align}
{\rm det} \left[(y-\Omega)+\frac{g^2}{(2\pi)^3c}A^{+}(y)\right] = 0
\end{align}
are continuous functions of $g$ for $g \in[0,G_0]$ and hence the number of roots inside $D(\Omega,R_0)$ is constant for all $0 \le g \le G_0.$ Setting $g = 0$ we see that there are exactly $N.$ Moreover, from above it follows that for $g>0$ the sign of the imaginary part of each root cannot change.

Let $\lambda_j$ be the eigenvalues of $A^+(\Omega).$ For $g$ sufficiently small the $N$ roots $y_1(g),\dots,y_N(g)$ satisfy
\begin{align}
y_j = \Omega -\frac{g^2}{(2\pi)^3c} \lambda_j + O(g^2).
\end{align}
  In particular, since the imaginary part of $A^+(\Omega)$ is negative-definite it follows that $\Im{y_j} >0$ for $g>0$ sufficiently small. It follows by continuity of roots that for all $0<g<G_0,$ $y_1,\dots,y_N \in D(\Omega,R_0) \in \Omega_A$ and $\Im{y_j} >0,$ $j=1,\dots,N.$
\end{proof}
The following corollary follows immediately from the previous lemma and the fact that ${A^-(y)} = \overline{A^{+}(\overline{y})}.$ 
\begin{corollary}
Let $\Omega_A,$ $G_0,$ and $R_0$ be the same as in the previous lemma. Then
\begin{align}\label{eqn:f-_eqn}
( y -  \Omega) I + \frac{g^2}{(2\pi)^3 c} A^-(y) 
\end{align}
has exactly $N$ roots in the disk of radius $R_0$ centered at $\Omega,$ and no others in $\Omega_A.$ In particular, if $y_1,\dots,y_N$ are the roots from the previous lemma, then the roots of the determinant of (\ref{eqn:f-_eqn}) are $\bar{y_1},\dots,\bar{y}_N$. 
\end{corollary}

\begin{remark}
If $y = \lambda-i\Gamma$ $\lambda, \Gamma>0,$ is a pole of $H^+$ such that the dimension of the nullspace of $(H^{+})^{-1}(y)$ is the same as the algebraic multiplicity of the eigenvalue $0$ for $(H^{+})^{-1}(y)$, then a more explicit expression can be obtained for the corresponding residue. In particular, suppose that the nullspace of $(H^+)^{-1}$ is $d$-dimensional, and let  $\vec{u}_{j}(\xi)$ and $\vec{v}_{j}(\xi)$, $j=1,2,\ldots d$ denote the left and right unit eigenvectors (respectively) of $H^+(\xi),$ $|z-\xi|<\varepsilon,$ such that:
\begin{enumerate}
\item $u_j^*(\xi) v_k(\xi)=\delta_{j,k}$
\item $\lim_{y \to 0} u_j^*(y)(H^+)^{-1}(y)  =0 = \lim_{y \to 0} (H^+)^{-1}(y) v_j(y),$ $j=1,\dots,d.$
\end{enumerate} 
With these assumptions, the residue is given by
\begin{equation}
S(z) = \sum_{j=1}^{d} \frac{u_{j}(z) v_{j}(z)^{\star}}{1 + \frac{g^2}{(2\pi)^3 c} \frac{u_{j}^{\star}(z) (A^{+})'(z)v_{j}(z)}{u_{j}^{\star}(z)v_{j}(z)}} \, .
\end{equation}
The above formula can be derived from the results in~\cite{schumacher1985residue} for example.

For the case of non-simple poles, i.e. when $(H^{+})^{-1}(z)$ has a non-trivial Jordan block corresponding to the eigenvalue $0$,  expressions for the total residue $S(z)$ can be derived (see~\cite{gohberg1971operator} for example).
\end{remark}

Let $z_{j} = \lambda_{j} - i \Gamma_{j}$, $j=1,2,\ldots N$ denote the location of the poles of $H^{+}(y) - H^{-}(y)$ and $S(z_{j})$ denote the corresponding residues. Deforming the contour of integration in~\cref{eq:betafin} to $\theta = -\pi/6$, the solution $\vec{\beta}$ is given by
\begin{equation}
\begin{aligned}
\label{eq:betafin2}
\vec{\beta}(t) = \frac{e^{i \Omega t}}{2\pi i} &\int_{0}^{\infty e^{-i\pi/6}} e^{-izt} \left(  \left( z - \Omega + \frac{g^2}{(2\pi)^3 c} A^{+}(z) \right)^{-1} - \left( z - \Omega + \frac{g^2}{(2\pi)^3 c} A^{-}(z) \right)^{-1} \right) \vec{\beta}(0) \, dz \\
&+  \sum_{j=1}^{n_{p}} R(p_{j}) \vec{\beta}(0) e^{i(p_{j} + \Omega) t}
  + \sum_{j=1}^{N} S(z_{j}) \vec{\beta}(0) e^{-\Gamma_{j} t +
  i(-\lambda_{j} + \Omega) t}.
\end{aligned}
\end{equation}

In the following, we derive asymptotic formulae for the integrals appearing in the previous expression for $\vec{\beta}(t)$. Let $A_{0} = \lim_{z\to 0} A^{\pm}(z)$. Note that it follows from the definitions of $A^{\pm}(z)$ that their limiting values at the origin are identical. Suppose further that $g\leq g_0$ with $g_{0}$ sufficiently small so that $(\Omega - \gamma A_{0})$ is invertible, where $\gamma = \frac{g^2}{(2\pi)^3c}$. Finally, let $\phi_{1}, \phi_{2}$ be a partition of unity of $e^{-i\pi/6}(0,\infty)$, where $\phi_{1}$ is compactly supported and $1$ in the vicinity of the origin and $\phi_{2}$ is $0$ in the vicinity of the origin. In particular, $\phi_1(z)+\phi_2(z) = 1$ for all $z \in e^{-i\pi/6}(0,\infty).$
We write the integral in (\ref{eq:betafin2}) as the sum of the following two terms
\begin{equation}
\begin{aligned}
\vec{\alpha}_{1}(t) &= \frac{e^{i \Omega t}}{2\pi i} \int_{0}^{\infty e^{-i\pi/6}} e^{-izt}  \phi_{1}(z) \left(  \left( z - \Omega + \gamma A^{+}(z) \right)^{-1} -  \left( z - \Omega + \gamma A^{-}(z) \right)^{-1} \right) \vec{\beta}(0) \, dz \\
\vec{\alpha}_{2}(t) &= \frac{e^{i \Omega t}}{2\pi i} \int_{0}^{\infty e^{-i\pi/6}} e^{-izt}  \phi_{2}(z) \left(  \left( z - \Omega + \gamma A^{+}(z) \right)^{-1} -  \left( z - \Omega + \gamma A^{-}(z) \right)^{-1} \right) \vec{\beta}(0) \, dz 
\end{aligned}
\end{equation}
Here the support of $\phi_1,$ $z \in e^{-i\pi/6}(0,\delta),$ is chosen
so that $\|\left(z+\gamma (A^\pm(z) -A_{0}) \right) \left(\Omega - \gamma A_{0} \right)^{-1}\| <1$ on
$e^{-i\pi/6}(0,\delta)$. Moreover, the same value of $\delta$ can be chosen for all $g\leq g_{0}$.
In the following lemma, we derive an asymptotic expression for $\vec{\alpha}_{1}(t)$.
\begin{lemma}
For $g$ sufficiently small, the asymptotic expansion for
  $\vec{\alpha}_{1}(t)$ as $t \to \infty$ is given by
\begin{equation}
\vec{\alpha}_{1}(t) = \frac{g^2  e^{i \Omega t}}{2\pi^2 i c^2 t^3 \Omega^2} \vec{\beta}(0) + O\left(\frac{g^4}{t^3}\right)\vec{\beta}(0) + o\left( \frac{1}{t^3}\right)\vec{\beta}(0)
\end{equation}
\end{lemma}
\begin{proof}
Observe that as $z\to 0$, $A^{\pm}(z)$ have the following asymptotic behavior in the vicinity of the origin
\begin{equation}
A^{\pm}(z) = A_{0} + A_{1} z + \tilde{A}_{2} z^2 \log{|z|} + \left(\mp i\frac{4\pi^2}{c^2} + A_{2} \right) z^2 + O(|z|^3) \, ,
\end{equation}
where $A_{0},A_{1}, A_{2}, \tilde{A}_{2} \in \mathbb{C}^{N\times N}$. Given the fact that $A^{\pm}(z)$ are bounded at the origin and that $\gamma$ is also small, we can compute the inverses of $(z-\Omega + \gamma A^{\pm}(z))$ as a Neumann series given by 
\begin{equation}
(z-\Omega + \gamma A^{\pm}(z))^{-1} = -\left(\Omega - \gamma A_{0} \right)^{-1} \sum_{j=0}^{\infty}\left( z+ \gamma(A^{\pm}(z)-A_{0}) \right)^{j}\left( \Omega - \gamma A_{0} \right)^{-j}.
\end{equation}
This immediately implies
\begin{equation}
(z-\Omega + \gamma A^{+}(z))^{-1} - (z-\Omega + \gamma A^{-}(z))^{-1} = \frac{i\gamma 8\pi^2  z^2}{c^2}\left(\Omega -\gamma A_{0} \right)^{-2} + o(|z|^2) \, .
\end{equation}
The above result combined with the observation that
\begin{equation}
\int_{0}^{\infty e^{-i\pi/6}} z^2 \phi_{1}(z) e^{-izt} \, {\rm d}z = \frac{2i}{t^3} + O\left(\frac{1}{t^4}\right) \, \quad \, t\to \infty \, ,
\end{equation}
yields the following asymptotic expansion for $\vec{\alpha}_{1}(t)$
\begin{equation}
\begin{aligned}\label{eqn:alpha1}
\vec{\alpha}_{1}(t) &= -\frac{16\pi^2 \gamma e^{i \Omega t}}{2\pi i c^2 t^3} \left( \Omega - \gamma A_{0}\right)^{-2} \vec{\beta}(0) + o\left( \frac{1}{t^3}\right)\vec{\beta}(0) \, ,\\
&= -\frac{g^2  e^{i \Omega t}}{\pi^2 i c^2 t^3 \Omega^2} \vec{\beta}(0) + O\left(\frac{g^4}{t^3}\right)\vec{\beta}(0) + o\left( \frac{1}{t^3}\right)\vec{\beta}(0) \, .
\end{aligned}
\end{equation}
\end{proof}

In the following lemma, we derive an asymptotic expression for $\vec{\alpha}_{2}(t)$.
\begin{lemma}
Let $F(y)$ denote the matrix $f(y/c, r_{j,\ell})$, $j,\ell=1,2,\ldots N$, let $\rmax = \max_{j,\ell} r_{j,\ell}$ and let 
$$ C_{F} = \sup_{\textrm{Arg}(y) = -\pi/6} e^{-\frac{y \rmax}{c}} |F(y)| < \infty \, .$$
The solution $\vec{\alpha}_{2}(t)$ satisfies
\begin{equation}
\vec{\alpha}_{2}(t) = O\left(\frac{4 g^2 C_{F}e^{-\delta\frac{(ct-2\rmax)}{2c}}}{\pi^3 \Omega^2 (ct-2\rmax)^3 \sin^2(\pi/8)}(2c^2 + \delta(ct -2\rmax) (2c + \delta(ct - 2\rmax)) ) \right) \vec{\beta}(0)
\end{equation}
as $t \to \infty$.
\end{lemma}
\begin{proof}
Note that
\begin{equation}
\begin{aligned}
\vec{\alpha}_{2}(t) &= \frac{e^{i \Omega t}}{2\pi i} \int_{0}^{\infty e^{-i\pi/6}} e^{-izt}  \phi_{2}(z) \left(  \left( z - \Omega + \gamma A^{+}(z) \right)^{-1} -  \left( z - \Omega + \gamma A^{-}(z) \right)^{-1} \right) \vec{\beta}(0) \, {\rm d}z  \\
&= \frac{e^{i \Omega t - i\pi/6}}{2\pi i} \int_{0}^{\infty} e^{-t \left(iy \cos{\left(\pi/6\right)} + y \sin{\left(\pi/6\right)}\right)}   \phi_{2}(ye^{-i\pi/6}) \bigg(  \left( ye^{-i\pi/6} - \Omega + \gamma A^{+}(ye^{-i\pi/6}) \right)^{-1} -  \\
&\quad \quad \quad \left( ye^{-i\pi/6} - \Omega + \gamma A^{-}(ye^{-i\pi/6}) \right)^{-1} \bigg) \vec{\beta}(0) \, {\rm d}y  \\
&= \frac{e^{i \Omega t - i\pi/6}}{2\pi i} \int_{0}^{\infty} e^{-t \left(iy \cos{\left(\pi/6\right)} + y \sin{\left(\pi/6\right)}\right)}   \phi_{2}(ye^{-i\pi/6})  \left( ye^{-i\pi/6} - \Omega + \gamma A^{+}(ye^{-i\pi/6}) \right)^{-1}  \\
&\quad \quad \quad \gamma \left(A^{-}(ye^{-i\pi/6}) - A^{+}(ye^{-i\pi/6}) \right) \left( ye^{-i\pi/6} - \Omega + \gamma A^{-}(ye^{-i\pi/6}) \right)^{-1} \vec{\beta}(0) \, {\rm d}y
\end{aligned}
\end{equation}
Since $\phi_{2}(ye^{-i\pi/6})$ is zero in an interval close to the origin, the matrix inverses 
$$\left( ye^{-i\pi/6} - \Omega + \gamma A^{\pm}(ye^{-i\pi/6}) \right)^{-1}$$ are bounded by  $2/(\Omega \sin(\pi/8))$  (which follows from~\cref{eq:matinvbound}), 
and $$|A^{-} (ye^{-i\pi/6}) - A^{+} (ye^{-i\pi/6})| = 2\pi |y|^2 |e^{-y^2 R^2/c^2}| |F(ye^{-i\pi/6})| \leq 2\pi C_{F} e^{\frac{y\rmax}{c}} \,  .$$ 
Inserting these bounds into the expression for $\vec{\alpha}_2,$ we obtain
\begin{equation}
\begin{aligned}
|\vec{\alpha}_{2}(t)| &\leq  \frac{4C_{F}\gamma|\vec{\beta}(0)|}{\Omega^2 \sin^2(\pi/8)} \int_{\delta}^{\infty} y^2 e^{-yt/2} e^{ y r_{\textrm{max}}/c} \, dy \\
& =\frac{4 g^2 C_{F}e^{-\delta\frac{(ct-2\rmax)}{2c}}|\vec{\beta}(0)|}{\pi^3 \Omega^2 (ct-2\rmax)^3 \sin^2(\pi/8)}(2c^2 + \delta(ct -2\rmax) (2c + \delta(ct - 2\rmax)) ) \\
&= O\left(\frac{g^2 \delta^2 e^{-\delta t/2} |\vec{\beta}(0)|}{\Omega^2 ct}\right) \, .
\end{aligned}
\end{equation}
\end{proof}

\section{The continuum limit}

We now consider the continuum limit, in which the number of atoms is
taken to infinity with $R,$ $c,$ and $\Omega$ held fixed. To that end,
we suppose $g = \tilde{g}/\sqrt{N}$, and that the atoms are distributed according to some continuous density $\rho.$ We also assume that at any fixed time the probability amplitudes of the atoms are a continuous functions of their positions. With some abuse of notation, we denote the corresponding probability amplitude by $\beta(\bx,t).$ Taking the limit as $N\to \infty,$ we find that
\begin{align}
\frac{\partial}{\partial t} \beta(\bx,t) = -\frac{\tilde{g}^2}{(2\pi)^3}
  \int_{\mathbb{R}^3} \int_0^t \,\rho(\by) \beta(\by,\tau)e^{i\Omega (t-\tau)} \int_0^\infty k^2 e^{-ic|k|(t-\tau)-k^2R^2} f(k,|\bx-\by|)\,{\rm d}k\,{\rm d}\tau \, {\rm d}\by.
\end{align}
In the following, it will be convenient to rescale $\beta,$ defining $\tilde{\beta}$ by $\tilde{\beta}(\bx,t) = \sqrt{\rho (\bx)} \beta(\bx,t).$ Then
\begin{align}\label{eqn:beta_inf}
\frac{\partial}{\partial t} \tilde{\beta}(\bx,t) = -\frac{\tilde{g}^2}{(2\pi)^3} \int_{\mathbb{R}^3} \int_0^t\,\tilde{\beta}(\by,\tau)e^{i\Omega (t-\tau)} \int_0^\infty k^2 e^{-ic|k|(t-\tau)-k^2R^2} \sqrt{\rho(\bx)\,\rho(\by)} f(k,|\bx-\by|)\,{\rm d}k\,{\rm d}\tau\, {\rm d}\by.
\end{align}
Assuming $\rho$ to be compactly supported, it is straightforward to show that the operator $\mathcal{F}_k:L^2(\mathbb{R}^3) \to L^2(\mathbb{R}^3)$ defined by
$$\mathcal{F}_k[g](\bx) = \int_{\mathbb{R}^3} f(k,|\bx-\by|) g(\by) \sqrt{\rho(\bx) \rho(\by)}\,{\rm d}\by$$
is compact, symmetric, and positive semi-definite. Moreover, the family
of operators indexed by $k$ is uniformly bounded on $\mathbb{R}$ and
uniformly bounded on any compact subset of $\mathbb{C}$. Additionally, if $h:\mathbb{R}^+ \to \mathbb{R}^+$
with $h \in L^1$, and $h$ is not identically zero, then
$$\int_0^\infty h(k)\,\mathcal{F}_k\,{\rm d}k$$
is positive semi-definite on $L^2({\rm supp}(\rho)).$ 

As in the case in which the number of atoms is finite, we can take the
Laplace transform of (\ref{eqn:beta_inf}), yielding
\begin{align}\label{eqn:lap_B}
sB(\bx,s) - B(\bx,0) = -\frac{\tilde{g}^2}{(2\pi)^3}\Gamma_{s-i\Omega}[B](\bx,s).
\end{align}
Here $\Gamma_{s},$ $s \in \mathbb{C},$ is the operator defined by
$$\Gamma_s[g](\bx) = \int_0^\infty {\rm d}t \int_0^\infty {\rm d}k\, k^2
e^{-ickt-st-k^2R^2} \mathcal{F}_k[g](\bx) = \frac{1}{ic}\int_0^\infty {\rm d}k \,k^2  \frac{e^{-k^2R^2}}{\frac{s}{ic}+k} \mathcal{F}_k[g](\bx).$$
Inverting the Laplace transform, the solution $\tilde{\beta}(\bx,t)$ is given by
\begin{equation}
\begin{aligned}
\tilde{\beta}(\bx,t) &= \frac{1}{2\pi i} \int_{\sigma -i\infty}^{\sigma + i\infty} e^{st} \left( sI + \frac{\tilde{g}^2}{(2\pi)^3} \Gamma_{s-i\Omega} \right)^{-1} [\tilde{\beta}_{0}](\bx) \, ds \\ 
&= \frac{e^{i\Omega t}}{2\pi i} \int_{\sigma -i\infty}^{\sigma + i\infty} e^{st} \left( sI + i\Omega I+ \frac{\tilde{g}^2}{(2\pi)^3} \Gamma_{s} \right)^{-1} [\tilde{\beta}_{0}](\bx) \, ds \,,
\end{aligned}
\end{equation}
where $\sigma$ is a sufficiently large positive real number, and $\tilde{\beta}_{0} = \tilde{\beta}(\bx,0)$.

Note that $\Gamma_s$ is compact for all $s \in \mathbb{C} \setminus L.$
We summarize several of its properties in the following lemma. The proof
is almost identical to that for the finite-dimensional case. 

\begin{lemma}
$\Gamma_s$ is an analytic operator-valued function of $s \in
  \mathbb{C}$, except for $s$ on the negative imaginary axis, where it has a branch cut. Furthermore, if $0<y,$
$$ic \lim_{\epsilon \to 0^+} \Gamma_{-iy\pm \epsilon} =  \left(\Theta(y)\mathcal{F}_{y/c} \pm i\pi \frac{y^2}{c^2} e^{-y^2R^2/c^2} \mathcal{F}_{y/c} + \mathcal{G}_y \right),$$
where $\Theta$ is analytic in the right half-plane and $\mathcal{G}_y$
  is an entire operator-valued function of $y.$ 
  Here $\Theta(y)$ is as defined in~\cref{lem:Gammadef}, and
  \begin{equation}
  \mathcal{G}_{y} = \int_{0}^{\infty} \frac{k^2 e^{-k^2 R^2}}{k-y/c}\left(\mathcal{F}_{k} - \mathcal{F}_{y/c} \right) {\rm d}k \, .  
  \end{equation}
\end{lemma}

An immediate consequence of the above lemma is that the family of operators $\Gamma_{s}$ are uniformly bounded for $s\in\mathbb{C}$. Let $C_{\Gamma} = \sup_{y\in\mathbb{C}} |\Gamma_{y} |$.

\begin{figure}[h!]
\centering
\includegraphics[width=0.7\textwidth]{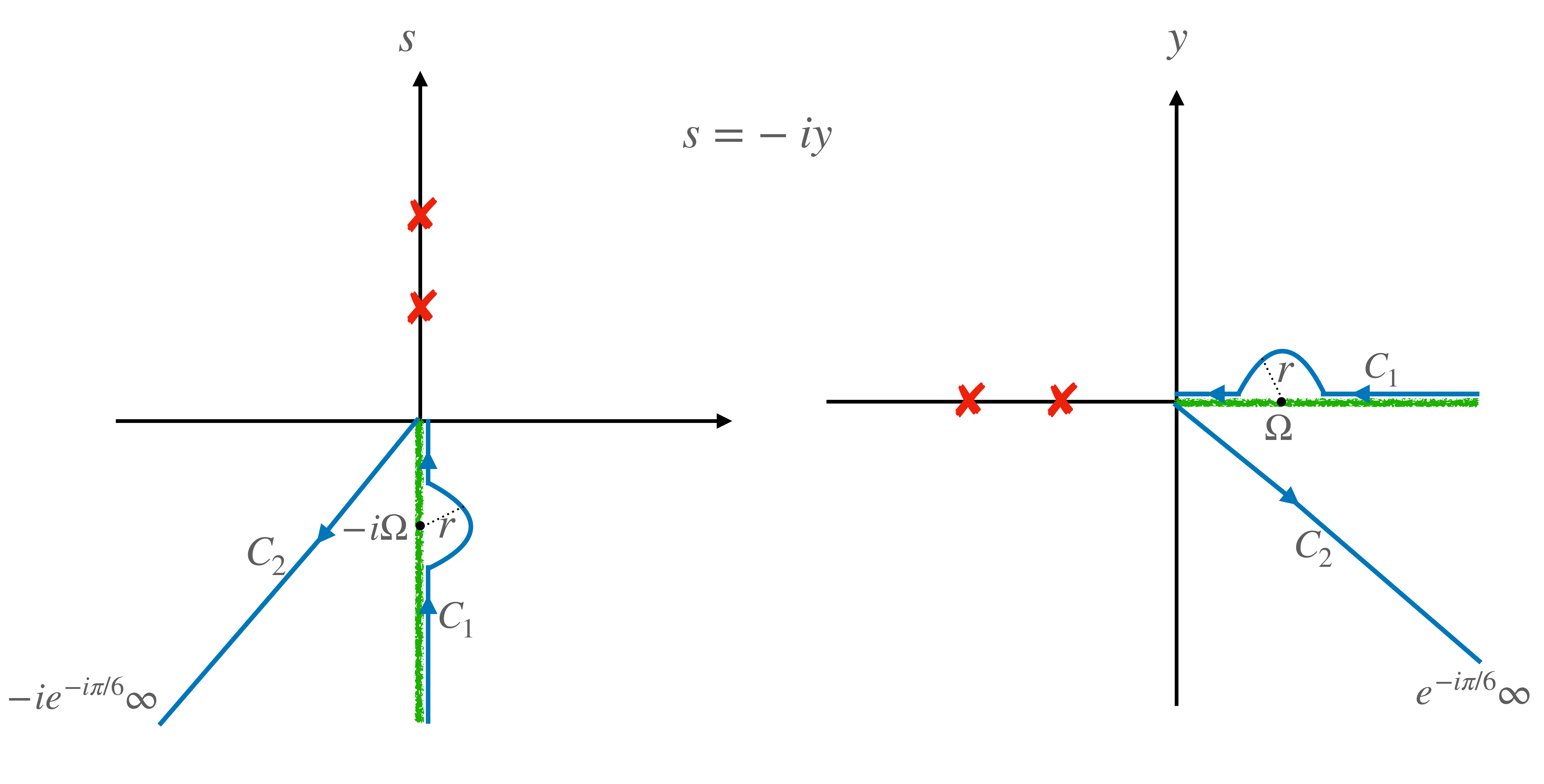}
  \caption{(left) The deformed contour $C = C_{1} \cup C_{2}$ for the inverse Laplace transform
  (blue), the poles of the integrand on the positive imaginary axis $p_{j}$
  (red), and the branch cut of the integrand, lying along the negative
  imaginary axis (green). (right) Same figure in the y-variable with $s=-iy$}\label{fig:contour-continuous}
\end{figure}

The following lemma is an operator version of Lemma \ref{lem:gam_mero}. The proof is almost identical to that
for the matrix-valued case, and follows directly from standard results in Fredholm theory.
\begin{lemma}
\label{lem:gamma-pole-cont}
The operator-valued function $\left(sI+i\Omega I
  +\frac{\tilde{g}^2}{(2\pi)^3}\Gamma_s\right)^{-1}$ is meromorphic on
  $\mathbb{C}\setminus L$ and its poles are on the positive imaginary axis. Here $L$ as before denotes the negative imaginary axis. Let $\lambda_j(y)$ denote the eigenvalues of $i\frac{\tilde{g}^2}{(2\pi)^3}\Gamma_{iy}$ and $v_j(y)$ the corresponding eigenfunctions. Then
  $$\left((y+\Omega)I-i\frac{\tilde{g}^2}{(2\pi)^3} \Gamma_{iy}\right)^{-1}$$
has a pole at $y_0$ if and only if $\lambda_j(y_0) = y_0+\Omega.$
  Moreover, the associated residue operator is
$$R_{y_0} = \frac{v_j(y_0)\,v^*_j(y_0)}{1-\lambda_j'(y_0)}.$$
Since $\Gamma_{iy}$ is compact for all $y>0$ and uniformly bounded, the number of poles is finite.
\end{lemma}
Let $p_1,\dots,p_{J}$ denote the poles described in the previous
lemma, and $R_1,\dots,R_J$ the corresponding residues. Furthermore, suppose that $\tilde{g} \leq g_{0}$ and $r = 2\frac{g_{0}^2 C_{\Gamma}}{(2\pi)^3}$. Then we note that for all $s$ such that $|s+i\Omega| \geq r$, and all $\tilde{g}\leq g_{0}$, the operator $((s+i\Omega)I + \frac{\tilde{g}^2}{(2\pi)^3} \Gamma_{s})$ has a bounded inverse, where the norm of the inverse is $O(1/|s|)$ as $s\to \infty$.
In this setting, we can deform the contour 
of integration of the inverse Laplace transform to the contour $C=C_{1} \cup C_{2}$ shown in~\cref{fig:contour-continuous} and the solution $\tilde{\beta}(\bx,t)$ is then given by
\begin{equation}
\begin{aligned}
\tilde{\beta}(\bx,t) &= \frac{e^{i\Omega t}}{2\pi i}\left( \int_{C_{1}} e^{st} \left(s+ i\Omega+ \frac{\tilde{g}^2}{(2\pi)^3} \Gamma_{s} \right)^{-1}\,[\tilde{\beta}_0](\bx)\,{\rm d} s + 
  \int_{C_{2}} e^{st}\left(s+ i\Omega + \frac{\tilde{g}^2}{(2\pi)^3} \Gamma_{s} \right)^{-1} [\tilde{\beta}_0](\bx)\,{\rm d} s \right) \,\\
&\quad\quad\quad-i \sum_{j=1}^J e^{i(p_j+\Omega)t} R_j [\tilde{\beta}_0](\bx) \, .
\end{aligned}
\end{equation}
Here if $s_{0} \subset C_{1}$ lies on the negative imaginary axis, the limiting value $\Gamma_{s_{0}}$ should be the limit of $\Gamma_{s}$ as $s\to s_{0}$ with $s$ in the fourth quadrant. 

Making the change of variable $s = -iy$ and in a slight abuse of notation while letting $C_{1}$ and $C_{2}$ denote the rotated contours in the $y-plane$ as well, the above expression for $\tilde{\beta}(\bx,t)$ can be rewritten as,
\begin{equation}
\label{eq:blapinv-cont}
\begin{aligned}
\tilde{\beta}(\bx,t) &= \frac{e^{i\Omega t}}{2\pi i}\bigg( \int_{C_{1}} e^{-iyt} \left(y- \Omega+ i\frac{\tilde{g}^2}{(2\pi)^3} \Gamma_{-iy} \right)^{-1}\,[\tilde{\beta}_0](\bx)\,{\rm d} y + \\
  &\quad \int_{C_{2}} e^{-iyt}\left(y- \Omega + i\frac{\tilde{g}^2}{(2\pi)^3} \Gamma_{-iy} \right)^{-1} [\tilde{\beta}_0](\bx)\,{\rm d} y \bigg) \,\\
&\quad\quad\quad-i \sum_{j=1}^J e^{i(p_j+\Omega)t} R_j [\tilde{\beta}_0](\bx) \,. 
\end{aligned}
\end{equation}
Moreover, note that $\Gamma_{-iy}$ now has a branch cut for $y$ on the positive real axis and the limiting value of $\Gamma_{y_{0}}$ for $y_{0} \subset C_{1}$ on the positive real axis, should be interpreted as the as limit of $\Gamma_{y}$ with $y\to y_{0}$ and  $y$ in the first quadrant.

We now turn to the analogue of the pole approximation. Let $\tilde{\Gamma}_{-iy}$ for $y$ in the fourth quadrant denote the analytical continuation onto the next Riemann sheet of the function $\Gamma_{-iy}$ for $y$ in the first quadrant, i.e.
\begin{equation}
ic \tilde{\Gamma}_{-iy} = \begin{cases}
ic \Gamma_{-iy} & \quad y \in \{ \Re{(y)}>0 \cap \Im(y)>0 \} \, , \\
ic \Gamma_{-iy} + 2\pi i \frac{y^2}{c^2} e^{-y^2 R^2/c^2} \mathcal{F}_{y/c} & \quad y \in \{ \Re(y)>0 \cap \Im(y) < 0 \} \, .
\end{cases} 
\end{equation}
Firstly, the function $\tilde{\Gamma}_{-iy}$ is an analytic function for $y$ in the right half plane. Moreover,  from the boundedness of $\Gamma_{y}$, it also follows that $\tilde{\Gamma}_{-iy}$ is uniformly bounded in the sector $\textrm{Arg}{(y)} \in (-\pi/5, \pi/2)$. Let $C_{\tilde{\Gamma}} = \sup_{\textrm{Arg}(y) \in (-\pi/5,\pi/2)} \| \tilde{\Gamma}_{-iy}\|$. Finally, suppose that $r$ in the contour $C_{2}$ is given by $r = 2\frac{G_{0}^2 \max{(C_{\Gamma}, C_{\tilde{\Gamma}}})}{(2\pi)^3}$. 
Since $\tilde{\Gamma}_{-iy} = \Gamma_{-iy}$ for $y$ in the first quadrant, we note that
\begin{equation}
\label{eq:cont-equality}
\int_{C_{1}} e^{-iyt}\left(y- \Omega + i\frac{\tilde{g}^2}{(2\pi)^3} \Gamma_{-iy} \right)^{-1} [\tilde{\beta}_0](\bx)\,{\rm d} y = 
\int_{C_{1}} e^{-iyt}\left(y- \Omega + i\frac{\tilde{g}^2}{(2\pi)^3} \tilde{\Gamma}_{-iy} \right)^{-1} [\tilde{\beta}_0](\bx)\,{\rm d} y
\end{equation}
Applying Cauchy's integral theorem to the region given by the interior of the wedge defined by the curves $C_{1} \cup C_{2}$ and the exterior of the disc or radius $r$ centered at $\Omega$, we get that 
\begin{equation}
\label{eq:cont-deform-cont}
\begin{aligned}
\int_{C_{1}} e^{-iyt}\left(y- \Omega + i\frac{\tilde{g}^2}{(2\pi)^3} \tilde{\Gamma}_{-iy} \right)^{-1} [\tilde{\beta}_0](\bx)\,{\rm d} y &= 
- \int_{C_{2}} e^{-iyt}\left(y- \Omega + i\frac{\tilde{g}^2}{(2\pi)^3} \tilde{\Gamma}_{-iy} \right)^{-1} [\tilde{\beta}_0](\bx)\,{\rm d} y  \\
&- \int_{\partial B_{r}(\Omega)} e^{-iyt}\left(y- \Omega + i\frac{\tilde{g}^2}{(2\pi)^3} \tilde{\Gamma}_{-iy} \right)^{-1} [\tilde{\beta}_0](\bx)\,{\rm d} y \, .
\end{aligned}
\end{equation}
The pole approximation is essentially the sum of residues of the all the poles of the operator $\left(y- \Omega + i\frac{\tilde{g}^2}{(2\pi)^3} \tilde{\Gamma}_{-iy} \right)^{-1}$ contained in the disc $B_{r}(\Omega)$. In the following lemma, we characterize the poles of the operator contained in $B_{r}(\Omega)$.

\begin{lemma}
Suppose that $y_{0}$ is a pole of $\left(y- \Omega + i\frac{\tilde{g}^2}{(2\pi)^3} \tilde{\Gamma}_{-iy} \right)^{-1}$ contained in $B_{r}(\Omega)$, then $\Im(y_{0}) \leq 0$. Moreover, there exist countably many poles of $\left(y- \Omega + i\frac{\tilde{g}^2}{(2\pi)^3} \tilde{\Gamma}_{-iy} \right)^{-1}$ in $B_{r}(\Omega)$
with the only possible accumulation point of the poles of being $y=\Omega.$
\end{lemma}
\begin{proof}
The proof follows directly from the observation that the operator is compact and analytic on $B_{r}(\Omega).$ As such, the only accumulation point is at $y=\Omega$ (see \cite{kriegl2011denjoy,ribarivc1969analytic} for example). Moreover, the fact that there are no poles with $\Im(y) >0$ follows from the fact that $\tilde{\Gamma}_{-iy} = \Gamma_{iy}$ for $\Im(y)>0$, and~\cref{lem:gamma-pole-cont}.
\end{proof}

Let $z_{j}$, $j=1,2,\ldots J_{\varepsilon} < \infty$ denote the poles of the operator $\left(y- \Omega + i\frac{\tilde{g}^2}{(2\pi)^3} \tilde{\Gamma}_{-iy} \right)^{-1}$ in $B_{r}(\Omega) \setminus B_{\varepsilon}(\Omega)$, then from Cauchy's integral formula, we get
\begin{equation}
\label{eq:res-cont}
\begin{aligned}
- \int_{\partial B_{r}(\Omega)} &e^{-iyt} \left(y- \Omega + i\frac{\tilde{g}^2}{(2\pi)^3} \tilde{\Gamma}_{-iy} \right)^{-1} [\tilde{\beta}_0](\bx)\,{\rm d} y =  \\
& - \int_{\partial B_{\varepsilon}(\Omega)} e^{-iyt} \left(y- \Omega + i\frac{\tilde{g}^2}{(2\pi)^3} \tilde{\Gamma}_{-iy} \right)^{-1} [\tilde{\beta}_0](\bx)\,{\rm d} y + 
2 \pi i \sum_{j}^{J_{\varepsilon}} e^{-iz_{j}t} S_{j}[\tilde{\beta}_{0}](\bx,t) \, ,
\end{aligned}
\end{equation}
where the residues $S_{j}[\tilde{\beta}_{0}](\bx,t)$ are given by
\begin{equation}
S_{j}[\tilde{\beta}_{0}](\bx,t) = -\frac{e^{iz_{j}t}}{2\pi i} \lim_{\varepsilon \to 0^{+}} \int_{|\xi - z_{j}| = \varepsilon} e^{-iyt} \left(y- \Omega + i\frac{\tilde{g}^2}{(2\pi)^3} \tilde{\Gamma}_{-iy} \right)^{-1} [\tilde{\beta}_0](\bx)\,{\rm d} y \, .
\end{equation}

Combining~\cref{eq:blapinv-cont,eq:cont-equality,eq:cont-deform-cont,eq:res-cont}, we get the following expression for the solution
\begin{equation}
\label{eq:betafin22}
\begin{aligned}
\tilde{\beta}(\bx,t) &= \frac{e^{i\Omega t}}{2\pi i} \int_{0}^{e^{-i\pi/6} \infty} e^{-iyt} \bigg( \left(y- \Omega+ i\frac{\tilde{g}^2}{(2\pi)^3} \Gamma_{-iy} \right)^{-1}  - \left(y- \Omega + i\frac{\tilde{g}^2}{(2\pi)^3} \tilde{\Gamma}_{-iy} \right)^{-1}\bigg) [\tilde{\beta}_0](\bx)\,{\rm d} y  \,\\
&-\frac{e^{i\Omega t}}{2\pi i} \int_{\partial B_{\varepsilon}(\Omega)} e^{-iyt} \left(y- \Omega + i\frac{\tilde{g}^2}{(2\pi)^3} \tilde{\Gamma}_{-iy} \right)^{-1} [\tilde{\beta}_0](\bx)\,{\rm d} y  \\
&\quad\quad\quad-i \sum_{j=1}^J e^{i(p_j+\Omega)t} R_j [\tilde{\beta}_0](\bx) + \sum_{j}^{J_{\varepsilon}} e^{i(-z_{j}+\Omega)t} S_{j}[\tilde{\beta}_{0}](\bx,t) \, .
\end{aligned}
\end{equation}

We conclude with the following lemma, which establishes the large time
asymptotic behavior of the integral in \eqref{eq:betafin22}.  Let $\cA_{z}^{+} = ic \Gamma_{-iz}$, and $\cA_{z}^{-} = ic \tilde{\Gamma}_{-iz}$, and $\gamma = \tilde{g}^2/(2\pi)^3 c$. Moreover let $\cA_{0} = \lim_{z\to 0} \cA^{\pm}(z)$. Note that it follows from the definition of $\tilde{\Gamma}$ and $\Gamma$ that the limiting values of $\cA^{\pm}(z)$ at the origin are identical.
\begin{lemma}
Suppose that  $-\Omega I + \gamma \mathcal{A}_0$ has a bounded inverse, and 
$$\left\| \left(z+  \gamma (\mathcal{A}^{\pm}_z-\mathcal{A}_0) \right)(-\Omega I + \gamma \mathcal{A}_0)^{-1} \right\|<1$$
for all $|z| <\delta.$ Let $I(\bx,t)$ denote the integral over the ray $(0,\infty)e^{-i\pi/6}$ in (\ref{eq:betafin22}). Then
\begin{equation}
\begin{aligned}
I(\bx,t) &= -\frac{16\pi^2 \gamma e^{i\Omega t}}{2\pi i c^2 t^3} (\Omega - \gamma \mathcal{A}_0)^{-2}  [\tilde{\beta}_0](\bx) +o\left(\frac{1}{t^3}\right)\tilde{\beta}_{0}(\bx) +  O\left(\frac{g^2 \delta^2 e^{-\delta t/2}}{\Omega^2 ct}\right)\tilde{\beta}_{0}(\bx)  \, , \\
&=-\frac{g^2 e^{i\Omega t}}{\pi^2 i c^2 t^3 \Omega^2} \tilde{\beta}_0(\bx) + O\left(\frac{g^4}{t^3} \right) \tilde{\beta}_{0}(\bx) + o\left(\frac{1}{t^3}\right)\tilde{\beta}_{0}(\bx) +  O\left(\frac{g^2 \delta^2 e^{-\delta t/2}}{\Omega^2 ct}\right)\tilde{\beta}_{0}(\bx) \, .
\end{aligned}
\end{equation}
\end{lemma}
\begin{proof}
The proof is analogous to the proof in the finite dimensional case.
\end{proof}

\section{Conclusion}
We have presented a detailed asymptotic analysis of the system of integro-differential
equations~\eqref{eq:asystem} describing a collection of localized atoms
interacting with a photon field. In particular, we show that for weak
coupling strengths, the solution at short times is well-approximated by a sum of
decaying exponentials, for which the decay rates and corresponding Lamb
shifts are given by the poles of the determinant of an analytic
matrix-valued function.
This result is a refinement of the ``pole
approximation'' commonly used in the standard Wigner-Weisskopf theory of spontaneous emission.
At large times, the solution decays like $O(1/t^3)$, with
an explicit constant expressed in terms of the resonant
frequencies of the atoms and the coupling strength. For
strong coupling parameters, the solution is dominated by a sum of
oscillatory exponentials at all times. We also extend our analysis to the
continuum limit, in which the atoms are assumed to be distributed
according to a known density. 

In many practical settings, the solution is not dominated by
oscillatory exponentials, which suggests that there is an upper bound on the
coupling strength. When the oscillatory exponentials are
absent, the pole approximation and the long time $O(1/t^3)$
asymptotic decay are the dominant contributions to the
solution. This analysis therefore provides a reasonable approximation for a
system of atoms and the dynamics of the atomic amplitudes can be computed
in $O(N^4)$  operations independent of the final time horizon $T$, where
$N$ is the number of atoms in the system (see~\cref{rem:flopcount}). For
moderately-sized systems, say 
$N<1000$, for example, this approach may provide a good alternative to
the direct numerical solution of \eqref{eq:asystem}, which typically scales like $O(N^2 T^2)$ even
after using fast algorithms~\cite{hoskins2021fast}. A detailed
comparison of our asymptotics
to the numerical solution of~\eqref{eq:asystem} will be presented in an
forthcoming paper. 

Lastly, our analysis of
the continuum limit enables the calculation
of the physically relevant contributions of the pole
approximation to the system of partial differential equations governing
collective spontaneous emission of random and structured media described
in~\cite{kraisler21}. 

\bibliographystyle{ieeetr}
\bibliography{qo2}
\end{document}